# Abnormal wave propagation of high-*k* modes in tilted linear-crossing metamaterials

Zhiwei Guo[*], Haitao Jiang, and Hong Chen

*Key Laboratory of Advanced Micro-structure Materials, MOE, School of Physics Science and Engineering,*

*Tongji University, Shanghai 200092, China*

**Abstract**

The propagation properties of electromagnetic (EM) waves depend on media dispersion in momentum space, which can be characterized using iso-frequency contours (IFCs). Topological IFC transitions from a closed ellipsoid to an open hyperboloid have been widely studied and applied in the fields of imaging and long-range energy transfer. Normal linear-crossing metamaterials (NLCMMs), which undergo a new type of topological transition between dielectric-type and metal-type hyperbolic media, are attracting attention because of their potential as a new avenue for controlling light propagation and verifying unusual phenomena involving zero-index and hyperbolic media. In this work, we treat the rotation of the optical axis as a new degree of freedom and theoretically propose a tilted linear-crossing metamaterials (TLCMM). Specifically, the conical dispersions of the NLCMM and TLCMM have the same shapes as type-I and type-II Dirac cones, respectively, in condensed matter physics. Upon rotating the optical axis angle such that it is equal to the cone angle, we find that this special TLCMM has the shape of a type-III Dirac cone. This critical TLCMM can have many unique properties and is of a fundamentally different nature than neighboring phases. When EM waves with large wave vectors are incident to a metamaterial with an open IFC to free space, the incident EM wave is strongly reflected due to wave-vector mismatch. Here, we use boundary conditions and the causality law to reveal that TLCMM high-*k* modes can achieve abnormal refraction without reflection and filtering. Moreover, these phenomenon are observed experimentally in a planar circuit-based system. The circuit-based TLCMM not only provides a versatile platform for the study of robust negative refraction phenomena in metamaterials, but also has a planar structure that is easier to integrate. Our results regarding the manipulation of EM waves may enable their use in planar-integrated photonics including for directional propagation, cloaking, and switching.

## 1. Introduction

Anisotropic media exhibit some intriguing features when compared to their isotropic counterparts. For example, a hyperbolic metamaterial (HMM) exhibits an iso-frequency contour (IFC) that is an open hyperboloid and in which the principal components of its electric or magnetic tensor have opposite signs [1–6]. The topological transition of dispersion from a closed sphere in isotropic media to an open hyperboloid in anisotropic media has been achieved by various means, including changing the real [7–9] and imaginary parts [10, 11] of electromagnetic (EM) parameters. Like passive control, active control of the topological transition of dispersion using electrically controllable metamaterials is an important research topic [12–18]. Because of their open IFCs, HMMs support propagating electromagnetic (EM) waves with large wave vectors [19, 20]. Thus far, the capabilities of HMMs have been demonstrated in numerous applications that utilize their exotic high-$k$ modes, such as enhanced spontaneous emission [8], super-resolution imaging [21, 22], high-sensitivity sensors [23–25], long-range interactions [26–29], fingerprinting [30], wavefront manipulation [31], and optical pulling forces [32, 33].

A topological transition from a metal-type HMM to a dielectric-type HMM achieved by tuning the vertical EM-wave parameters of the anisotropic medium was proposed recently [34, 35]. This is different from the aforementioned topological transition that is achieved by controlling the in-plane anisotropic EM parameters in a two-dimensional (2D) case. Specifically, the critical IFC in this transition, in which the in-plane permeability signs are opposite while the permittivity tends to zero, corresponds to two intersecting lines and produces a normal linear-crossing metamaterial (NLCMM) [34, 35]. On one hand, the group velocity is perpendicular to the phase velocity because of the linear dispersion of the NLCMM. This leads to zero phase accumulation when the EM waves propagate in NLCMMs. This effect is similar to that noted in zero-index media. On the other hand, NLCMMs can support high-$k$ modes and can achieve all-angle negative refraction just like HMMs. Therefore, NLCMMs simultaneously possess the characteristics of HMMs and zero-index media. Moreover, NLCMMs can have many unique capabilities, such as directional propagation, beam splitting, and super-resolution imaging with partial cloaking [34]. Recently, NLCMMs have been proposed to generate non-diffracted Bessel beams with self-healing. This greatly improves the methods of producing Bessel beams. Using directional propagation and negative refraction of NLCMM, a point source can be used instead of a plane wave to produce a Bessel beam through a planar prism [35].

Very recently, manipulating the optical axis orientations of anisotropic media has emerged as a new way to manipulate EM waves. Examples include wide-angle enhanced optical absorption [36] and single-photon extraction [37]. In this work, we first note that the light-matter interaction can be controlled effectively by rotating the optical axes of the NLCMMs. Interestingly, the IFC of the tilted linear-crossing metamaterial (TLCMM) exhibits the shape of a type-II Dirac cone [38]. This allows TLCMMs to be used to control EM waves in new ways. In particular, when the rotation angle of the optical axis is equal to the cone angle of the NLCMM, the TLCMM has the shape of a type-III Dirac cone [39]. This critical TLCMM is of a fundamentally different nature than the neighboring phases and can be used to achieve robust negative refraction without reflection, just like topological Weyl crystals [40].

Using two dimensional (2D) transmission lines (TLs), we design and fabricate a circuit-based TLCMM that is intended to operate in a microwave regime. The effective permittivities and permeabilities of the circuit-based metamaterials can be flexibly tuned using the lumped elements in the TL system [41, 42]. Anisotropic artificial microstructures with magnetic responses [43, 44] and planarization [45–54] are presently attracting increased attention. Importantly, these circuit-based anisotropic magnetic media have subwavelength scale and planarization attributes. They provide a versatile platform for study of novel physical phenomena [55–57]. Using the near-field detection method, we demonstrate negative refraction without reflection in a circuit-based TLCMM. Although the effective permittivity of a TLCMM is close to zero, our research results are based on use of a rotating optical axis instead of doping to achieve transmission without reflection. This provides a new approach to achieving impedance matching of zero-index media [58]. In addition, the TLCMM is intended for use in the design of new anisotropic optical cavities [59, 60] and Moiré hyperbolic metasurfaces [61]. Our findings not only present a type of anisotropic metamaterial with special dispersion, but are also quite useful in a variety of applications related to planar integrated photonics, including directional propagation, cloaking, and filters.

## 2. Simulation and Discussion

The EM responses of materials depend on their permittivities and permeabilities. Thus far, most of the hyperbolic dispersions in artificial and natural optical media originate solely from the electric response [1–6], which restricts material functionality to one polarization of light [43]. Using the TL system, we design magnetic TLCMMs ($\varepsilon, \hat{\mu}$) and uncover their associated unique properties. The relative permeability tensor of the anisotropic magnetic medium is given by

$$\hat{\mu} = \begin{pmatrix} \mu_{xx} & 0 & 0 \\ 0 & \mu_{yy} & 0 \\ 0 & 0 & \mu_{zz} \end{pmatrix}. \tag{1}$$

We consider the uniaxial case, $\mu_{xx} = \mu_{yy} = \mu_\perp$ and $\mu_{zz} = \mu_{//}$. The subscripts $\perp$ and $//$ indicate that the components are perpendicular and parallel, respectively, to the optical axis, which is along the $z$-direction. In particular, the dispersion for the transverse electric (TE) polarization ($H_x, H_y, E_z$) can be written as

$$\frac{k_x^2 + k_y^2}{\mu_{//}} + \frac{k_z^2}{\mu_\perp} = \varepsilon k_0^2, \tag{2}$$

where $k_x$, $k_y$, and $k_z$ are the $x$, $y$, and $z$ components of the wave vector, respectively. $k_0$ is the wave vector in free space. A hyperbolic topological transition from the metal-type HMM to the dielectric-type HMM can be achieved by tuning the sign of $\varepsilon$ from negative to positive while keeping the signs of $\mu_{//}$ and $\mu_\perp$ positive and negative, respectively. Specifically, the transition point ($\varepsilon \to 0$) in this topological transition corresponds to the NLCMM with a conical dispersion structure. The corresponding 3D IFC is shown schematically in Fig. 1(a). Given tilted anisotropic media rotating at an angle $\alpha$ along the *y* axis, the relative permeability tensor in Eq. (1) should be rewritten as [62, 63]:

$$\hat{\mu}_T = \hat{T}^{-1}(\alpha)\hat{\mu}\hat{T}(\alpha),\qquad(3)$$

where the rotation matrix has the form

$$\hat{T}(\alpha) = \begin{pmatrix} \cos(\alpha) & 0 & -\sin(\alpha) \\ 0 & 1 & 0 \\ \sin(\alpha) & 0 & \cos(\alpha) \end{pmatrix}.\qquad(4)$$

In this case, the dispersion relation can be expressed as

$$Ak_x^2 + Bk_z^2 + Ck_xk_z + \mu_\perp k_y^2 = \varepsilon\mu_{//}\mu_\perp k_0^2,\qquad(5)$$

where $A = \mu_\perp \cos^2(\alpha) + \mu_{//}\sin^2(\alpha)$, $B = \mu_\perp \sin^2(\alpha) + \mu_{//}\cos^2(\alpha)$, and $C = \sin(2\alpha)(\mu_\perp - \mu_{//})$. Based on Eq. (5), the 3D IFC of the TLCMM ($\varepsilon \to 0, \mu_\perp = 1, \mu_{//} = -0.3$) is shown schematically in Fig. 1(b). For simplicity, we reduce the 3D anisotropic material to a 2D case in which the dimension in the *y*-direction is not considered. The 2D IFC of an NLCMM is shown in Fig. 1(c), where the directions of group velocities (energy flow density) are marked by green arrows. Because of the linear dispersion of the NLCMM, the direction of the refracted angle is fixed regardless of the incident angle. Under near-field excitation, the light in the media are collimated and propagate only in two fixed directions. After the optical axis rotates by angle $\alpha$, the 2D IFC is shown in Fig. 1(d). By comparing Figs. 1(c) and 1(d), one can find that the direction of the group velocity changes due to rotation of the optical axis, and this change significantly affects interactions between light and matter.

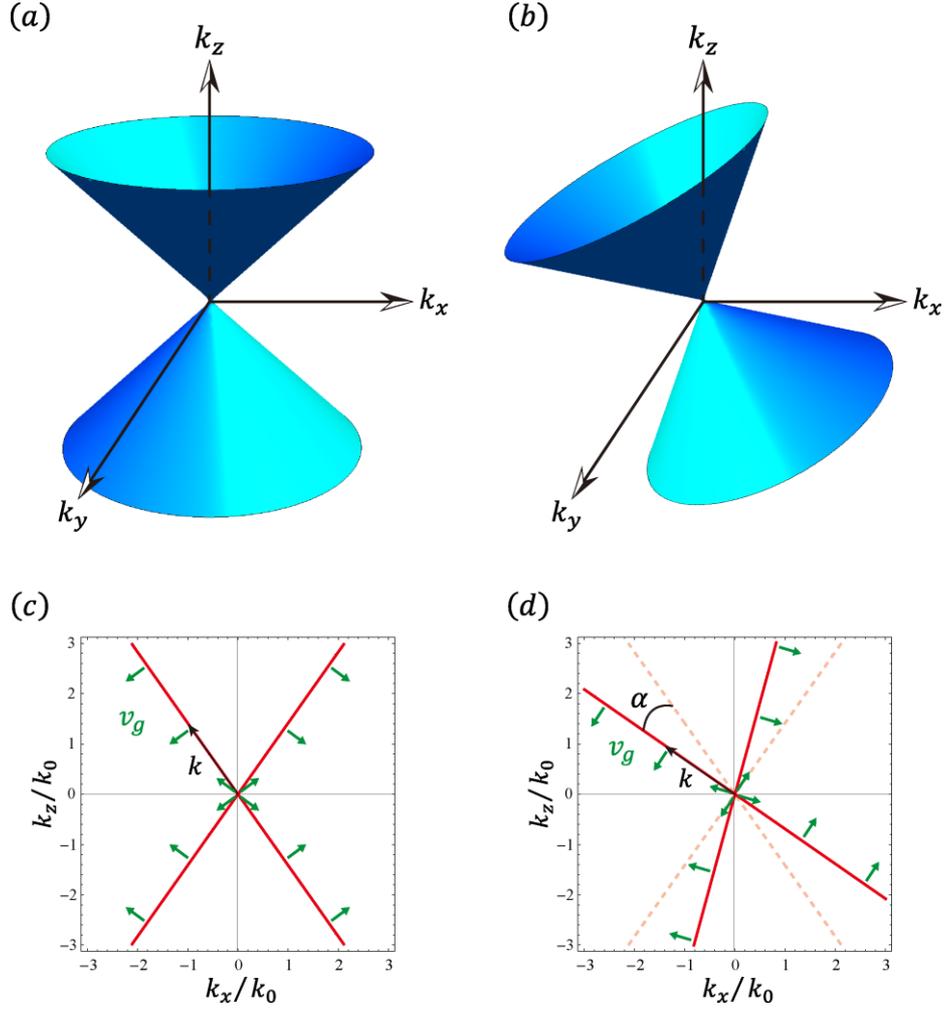

**FIG. 1. IFC changes during the transition from NLCMM to TLCMM when the anisotropic optical axis rotates at a particular angle.** (a) The 3D IFCs of the NLCMM with $\varepsilon \to 0$, $\mu_\perp = 1$, and $\mu_{//} = -0.3$ (b) Similar to (a) but for the TLCMM. (c) The cross graphs of (a) in the *xoz* plane. (d) Cross graphs of (b) in the *x-z* plane. For the TLCMM, the rotation angle of the optical axis is $\alpha$. The black and green arrows denote the group velocity and wave vector, respectively. The dashed contour in (d) corresponds to the NLCMM and is provided for comparison.

The transmission properties of EM waves change when the optical axis of the NLCMM is rotated. For example, after rotating the optical axis, the negative refraction phenomenon in a type-I LCMM can become positive refraction in a type-I TLCMM, and vice versa. In addition, when the rotation angle of the TLCMM is equal to its cone angle, its strong asymmetric linear dispersion causes abnormal refraction without reflection. To illustrate the above-mentioned points, we consider incident EM waves from two types of NLCMM and TLCMM to free space, as shown in Fig. 2. For the type-I NLCMM, the boundary conditions and the causality law allow us to determine that the incident EM waves (labeled using arrow 1) are negatively refracted (arrow 3) at the NLCMM ( $\mu_\perp > 0, \mu_{//} < 0,$ and $\varepsilon \to 0$ ) -air interface. This is accompanied by the reflected wave (arrow 2) shown in Fig. 2(a). The upper low is the IFC analysis, in which the IFCs expanding with the frequency is presented from the solid lines to dashed lines. Interestingly, when rotating the type-I NLCMM at an angle equal to the cone angle ( $\theta_c = art\tan(|\mu_\perp / \mu_{//}|^{1/2})$ ), one component of the linear dispersion of the type-I NLCMM is along the normal direction of the interface. As a result, the previous negative refraction becomes positive while the reflection is forbidden in the corresponding the type-I TLCMM, as shown in Fig. 2(b). Similarly, Figs. 2(c) and 2(d) show that, given normal positive refraction in the type-II NLCMM ( $\mu_\perp < 0, \mu_{//} > 0$ and $\varepsilon \to 0$), abnormal negative refraction without reflection occurs in the type-II TLCMM with special angle $\alpha = \theta_c$. On the whole, rotation of optical axis-induced EM-wave control is different from the previous methods and provides a new way to manipulate EM-wave propagation properties.

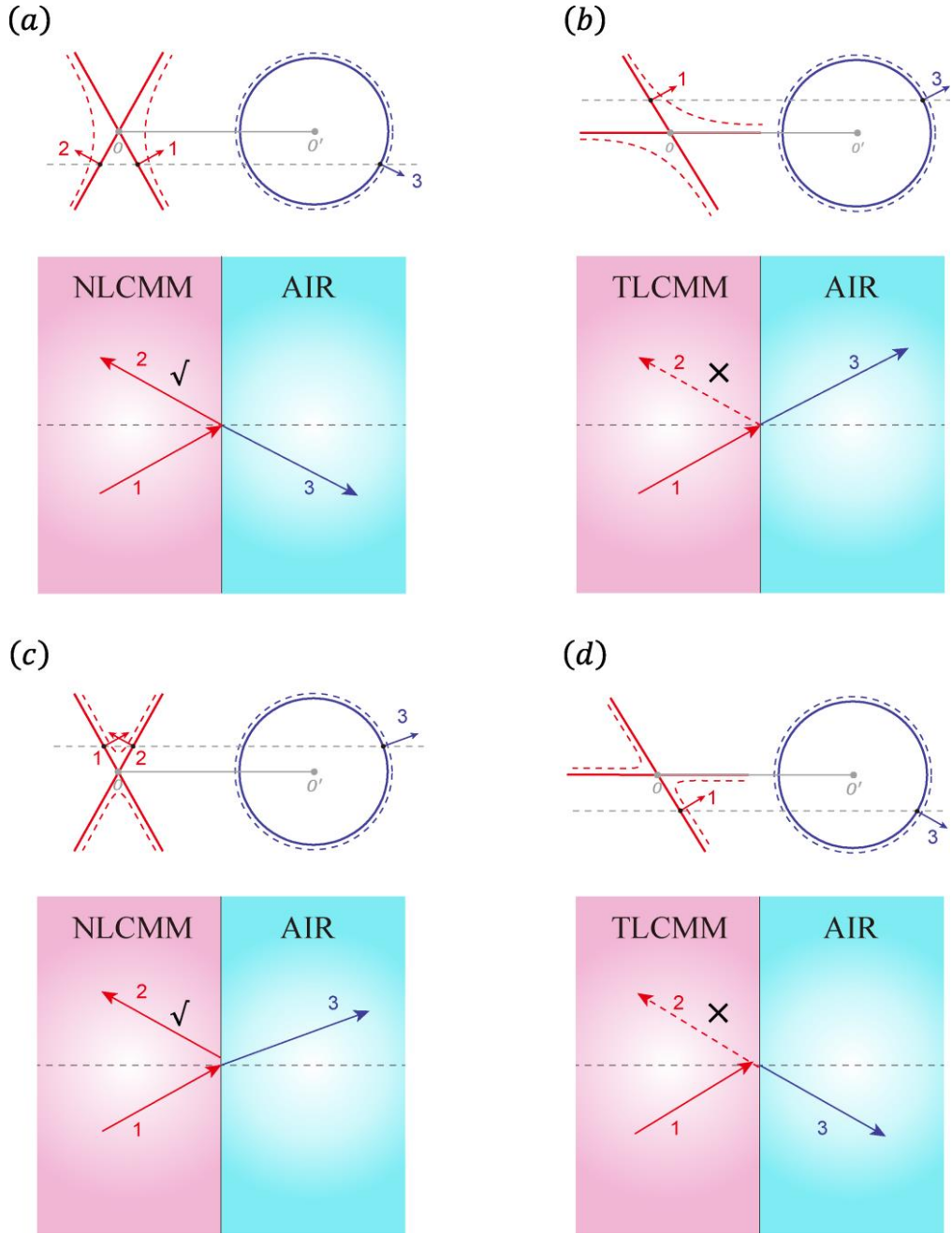

**FIG. 2. Schematics of TLCMM EM-wave responses.** (a) Negative refraction at the interface between the type-I NLCMM and air. (b) Positive refraction when the type-I NLCMM is rotated to become a type-I TLCMM. (c) Positive refraction at an interface between the type-II NLCMM and air. (d) Negative refraction when the type-II NLCMM is rotated to become a type-II TLCMM. The upper and lower rows are IFC analyses and associated beam propagations, respectively. The dashed lines denote the IFC that is calculated at the frequency slightly above the frequency of the solid lines. Arrows labeled 1, 2, and 3 indicate the incident, reflected, and refracted beams, respectively.

Under near-field excitation, high-$k$ modes with the largest densities of states are transmitted along fixed directions in anisotropic media using an open IFC [57]. Strong EM-wave reflection of the high-$k$ modes will occur at the NLCMM-air interface (see more details in the supplementary materials). It is this property that can be used to design hyperbolic waveguides and high-performance cavities. However, it can limit some applications related to EM-wave control in the far-field regime, such as antennas and information communications. This problem can be solved effectively by tuning the NLCMM to become a TLCMM. Strong reflections at the NLCMM-air interface can be eliminated when the optical axis of the NLCMM is rotated to the special angle $\alpha = \theta_c$ as previous IFC analysis in Fig. 2. Without loss of generality, two TLCMMs ($\theta_c \approx 61^o$) with $\alpha = 61^o$ and $\alpha = -61^o$ are used to demonstrate abnormal refraction in Fig. 3. For the type-I TLCMM ($\mu_\perp = -1, \mu_{//} = 0.3$) with $\alpha = 61^o$, directional emission under near-field excitation is shown in Fig. 3(a). Here, the directions of the high-$k$ modes are rotated because of optical axis rotation. In this case, when the high-$k$ modes in Fig. 3(a) impact the interface between the TLCMM and air, positive refraction occurs without reflection, as shown in Fig. 3(b). Specifically, the positive refraction becomes a negative refraction upon exchanging the signs of $\mu_\perp$ and $\mu_{//}$. For the type-II TLCMM ($\mu_\perp = 1, \mu_{//} = -0.3$) with $\alpha = 61^o$, abnormal negative refraction without reflection is shown in Fig. 3(c). Next, we consider the type-I TLCMM ($\mu_\perp = -1, \mu_{//} = 0.3$) with $\alpha = -61^o$. Directional emission under near-field excitation is shown in Fig. 3(d). When the high-$k$ modes in Fig. 3(d) impact the TLCMM-air interface, the high-$k$ mode incident to the interface refracts downward in the air, as shown in Fig. 3(e). In contrast, for the type-II TLCMM ($\mu_\perp = 1, \mu_{//} = -0.3$) with $\alpha = -61^o$, the high-$k$ mode incident on the interface refracts upward in the air, as shown in Fig. 3(f). Thus, the TLCMM emission patterns and the associated abnormal refraction are revealed by the numerical simulations in Fig. 3.

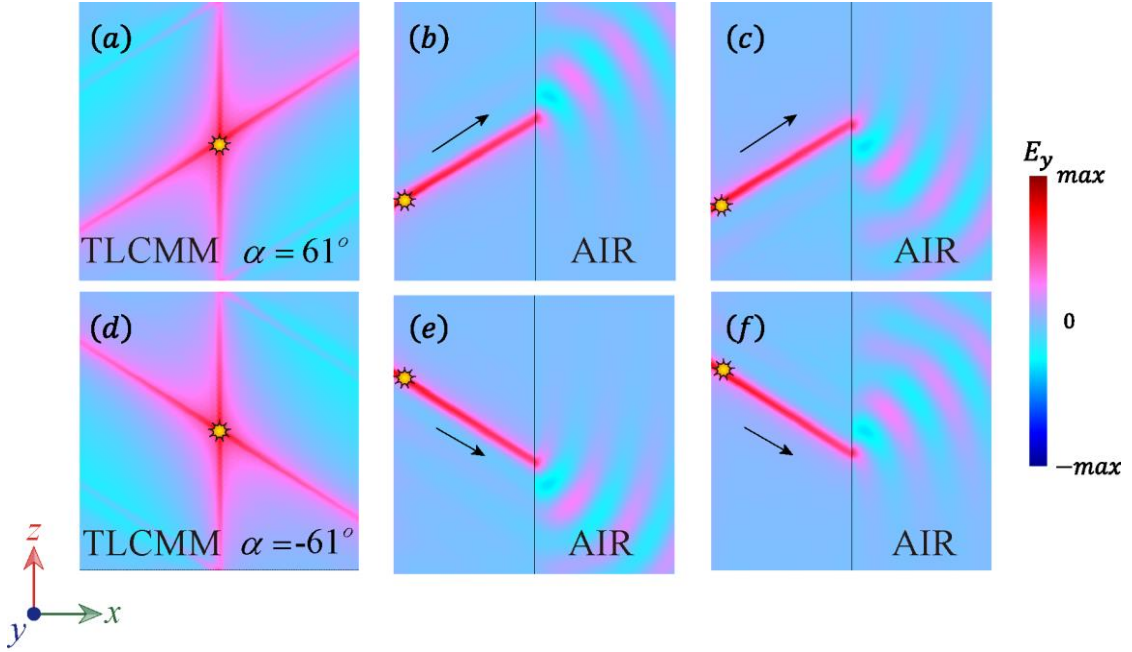

**FIG. 3. The $E_y$ patterns of high-$k$ modes in TLCMMs.** (a)–(c) Simulated electric field distribution $E_y$ in a TLCMM with $\alpha = 61^o$. (a) Directional emission in an NLCMM with $\mu_\perp = -1, \mu_{//} = 0.3$. A point source is placed at the center of the structure. (b) Positive refraction without reflection at the interface between a type-I TLCMM with $\mu_\perp = -1, \mu_{//} = 0.3$ and air. (c) Negative refraction without reflection at the interface between a type-II TLCMM with $\mu_\perp = 1, \mu_{//} = -0.3$ and air. (d)–(f) Similar to (a)–(c), but for the TLCMM with $\alpha = -61^o$. The type-I TLCMMs in (d) and (e) with $\mu_\perp = -1, \mu_{//} = 0.3$ and the type-II TLCMMs in (f) with $\mu_\perp = 1, \mu_{//} = -0.3$.

Although the TLCMM belongs to a reciprocal system, when the source is placed in the air instead of the TLCMM, refraction of the EM wave changes significantly when it passes through the air-TLCMM interface. This occurs because the large-$k$ mode cannot be excited when the source is placed in the free space. IFC analyses are shown for two types of NLCMM without rotation in Figs. 4(a) and 4(b). We can clearly see that the refracted angle is independent of the incident angle because of the linear dispersions of the NLCMMs. Moreover, exchanging the sign of $\mu_\perp$ or $\mu_{//}$ changes the corresponding refraction property. For example, for the NLCMM with $\mu_\perp = -1, \mu_{//} = 0.3$, the directional negative refraction can be determined from Fig. 4(a). However, the NLCMM with $\mu_\perp = 1, \mu_{//} = -0.3$ produces a directional positive refraction, as shown in Fig. 4(b). The two types of NLCMMs can be combined to demonstrate super-resolution with partial cloaking [34]. Next, EM-wave spatial filtering can be achieved using TLCMMs with $\alpha = -61^o$. The corresponding IFC analysis is shown in Figs. 4(c) and 4(d). For the type-II (type-I) TLCMM with $\mu_\perp = -1, \mu_{//} = 0.3$ ($\mu_\perp = 1, \mu_{//} = -0.3$), the EM waves are inhibited in the upper (lower) half of the structure, as shown in Fig. 4(e) (Fig. 4(f)). Because the refracted angle direction is fixed regardless of the incident angle, transmission is not influenced in TLCMMs even when a defect exists inside the shielding region (see more details in the supplementary materials).

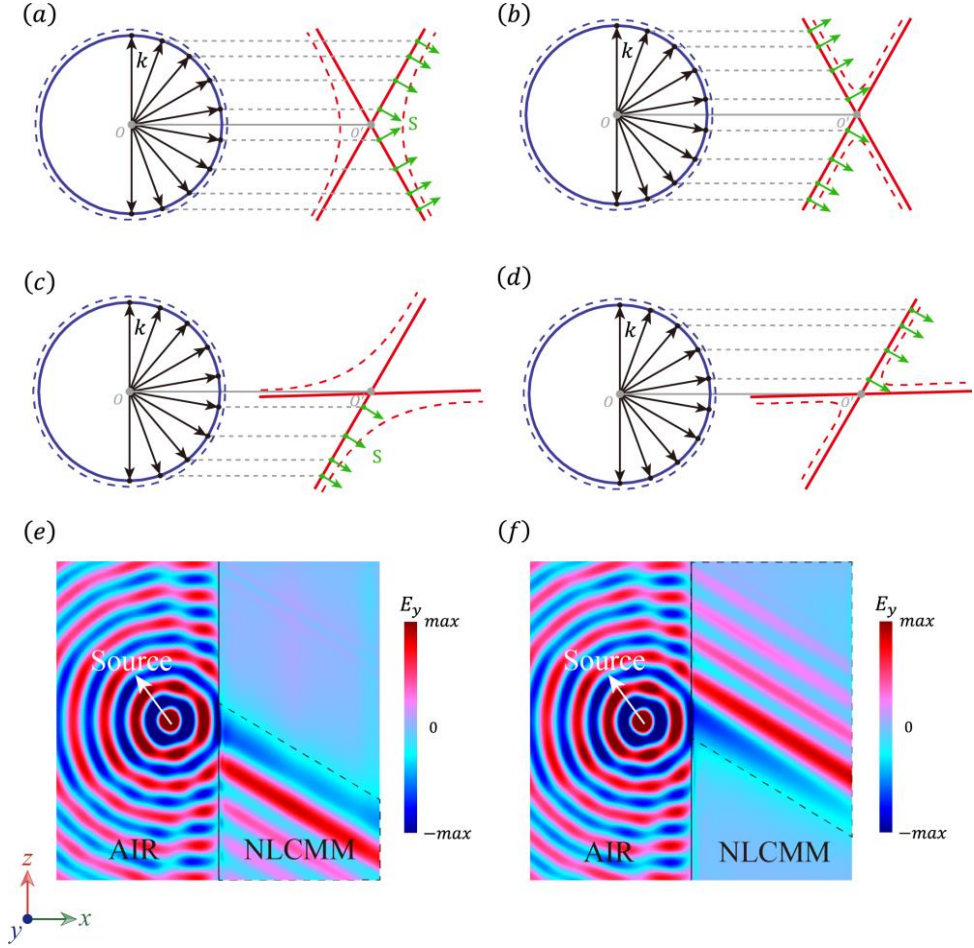

**FIG. 4. EM-wave spatial filtering using TLCMMs.** (a) (b) IFC analyses when EM waves with different incident angles impinge on two types of NLCMMs. (c) (d) Similar to (a) (b), but for two types of TLCMMs. The black and green arrows denote the group velocity and wave-vector, respectively. The type-II (type-I) TLCMM EM waves undergo directional propagation in the lower (upper) portion and are inhibited in the upper (lower) portion of the TLCMMs. (e) Simulated spatial filtering at the interface between air and the type-II TLCMM with $\mu_\perp = -1, \mu_{//} = 0.3$ and $\alpha = -61^o$. EM waves are inhibited in the upper half of the structure. (f) Similar to (e) but for the type-I TLCMM with $\mu_\perp = 1, \mu_{//} = -0.3$ and $\alpha = -61^o$. EM waves are inhibited in the lower half of the structure.

Anisotropic metamaterial heterojunctions provide an important way to achieve light control, such as concentration and beam splitting [64]. Here, we systematically study heterojunctions composed of two types of TLCMMs in Fig. 5. The schematic of the proposed setup is depicted in Figs. 5(a), 5(d), and 5(g). First, we consider an upward-pointing wedge interface between the two types of TLCMMs (Fig. 5(a)). When TLCMM1 and TLCMM2 are respectively characterized by $\mu_{\perp}=-1, \mu_{//}=0.3$ ($\alpha=61^o$) and $\mu_{\perp}=1, \mu_{//}=-0.3$ ($\alpha=61^o$), the EM waves are directional and refracted upward into the air, as shown in Fig. 5(b). However, when the EM parameters of the two types of TLCMMs are exchanged, the EM propagates only along the interface. Spatial filtering along the interface of TLCMMs is shown in Fig. 5(c). Second, given a straight interface between the two types of TLCMMs as in Fig. 5(d), directional transmission is achieved in Fig. 5(e). In the aforementioned example, TLCMM1 and TLCMM2 are respectively characterized by $\mu_{\perp}=1, \mu_{//}=-1$ ($\alpha=45^o$) and $\mu_{\perp}=-1, \mu_{//}=1$ ($\alpha=45^o$). Moreover, spatial filtering is also presented when the EM parameters of the two types of TLCMMs are exchanged (Fig. 5(f)). Finally, we consider a downward-pointing wedge interface between two types of TLCMMs in Fig. 5(g). When the anisotropic permeabilities are set to $\mu_{\perp}=1, \mu_{//}=-0.3$ ($\alpha=-61^o$) and $\mu_{\perp}=-1, \mu_{//}=0.3$ ($\alpha=-61^o$), directional propagation and spatial filtering are also demonstrated in Figs. 5(h) and 5(i), respectively. Based on the direction of the interface, one can flexibly choose two types of TLCMMs that can achieve the required directional transmission or spatial filtering.

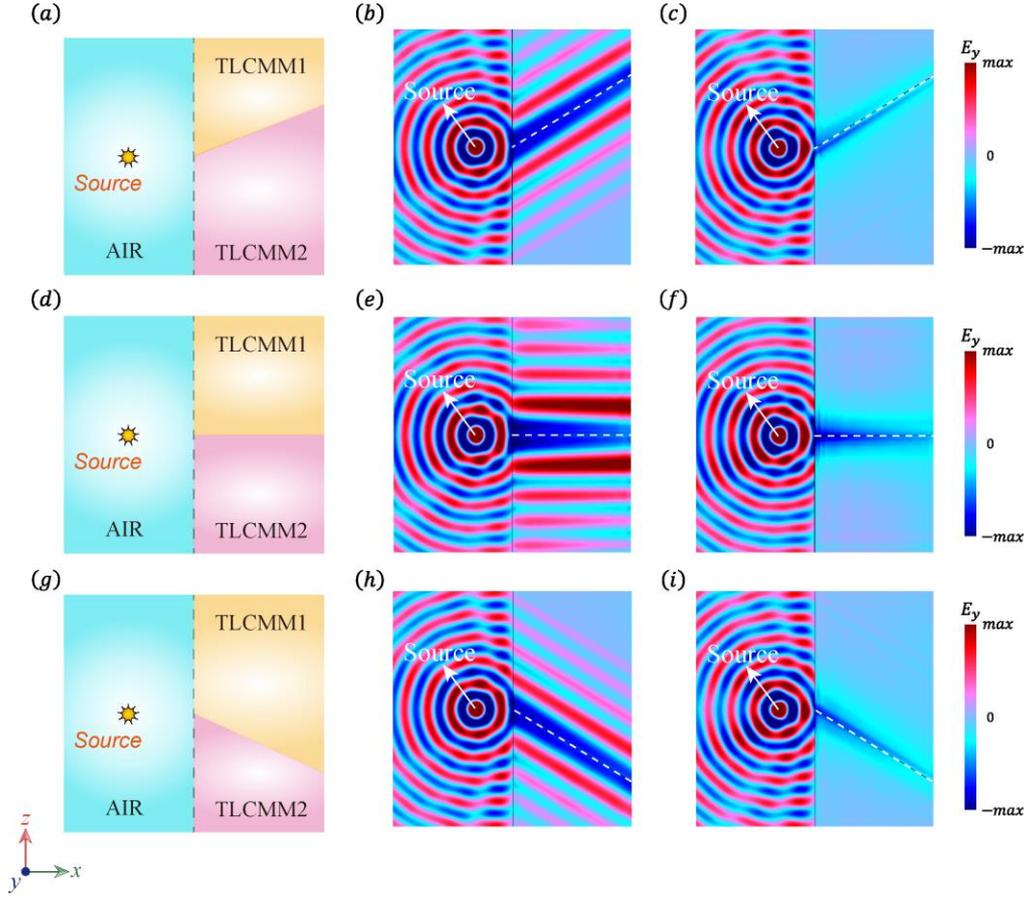

**FIG. 5. Directional transmission and spatial filtering achieved using a composite structure with two types of TLCMMs.** (a) Schematic diagram of the composite structure with an upward-pointing wedge interface between the two types of TLCMMs. (b) Obliquely upward directional transmission when TLCMM1 and TLCMM2 are characterized by $\mu_\perp = -1, \mu_{//} = 0.3$ ($\alpha = 61^o$) and $\mu_\perp = 1, \mu_{//} = -0.3$ ($\alpha = 61^o$), respectively. (c) Similar to (b), but the EM parameters of the two types of TLCMMs are exchanged. (d)–(f) Similar to (a)–(c) but for a straight interface between two types of TLCMMs. (e) Rightward directional transmission when TLCMM1 and TLCMM2 are characterized by $\mu_\perp = 1, \mu_{//} = -1$ ($\alpha = 45^o$) and $\mu_\perp = -1, \mu_{//} = 1$ ($\alpha = 45^o$), respectively. (f) Similar to (e), but the EM parameters of the two types of TLCMMs are exchanged. (g)–(i) Similar to (a)–(c) but for a downward-pointing wedge interface between two types of TLCMMs. (h) Obliquely downward directional transmission when TLCMM1 and TLCMM2 are characterized by $\mu_\perp = 1, \mu_{//} = -0.3$ ($\alpha = -61^o$) and $\mu_\perp = -1, \mu_{//} = 0.3$ ($\alpha = -61^o$), respectively. (i) Similar to (h), but the EM parameters of the two types of TLCMMs are exchanged.

## 3. Experimental Results

2D TLs loaded with lumped circuit elements can be used to achieve various permittivities and permeabilities [41, 42]. In this section, we use 2D TLs to design circuit-based TLCMMs for the microwave region. The samples fabricated on a F4B substrate with a thickness of $h = 1.6$ mm and relative permittivity of 2.2. The width of the microstrip line is $w = 2.8$ mm and the length of a unit cell is $d = 14$ mm. The structural factor of the designed circuit-based system is defined as $g = Z_0 / \eta_{eff} \approx 0.255$, where $Z_0$ and $\eta_{eff}$ are the characteristic impedance and effective wave impedance of the TL, respectively. In the proposed structure, an effective TLCMM is produced by loading series-lumped capacitors with $C = 3$ pF and shunted-lumped inductors with $L = 10$ nH. The corresponding effective circuit models for the normal double-positive (DPS) medium and TLCMM based on the TL system are shown in Figs. 6(a) and 6(b), respectively. For a quasi-static TE-polarized solution, we map voltage $v_g$ to $E_y$ by defining the potential difference, current $i_z$ to $H_x$ and current $i_x$ to $H_z$ via Ampere's law. Thus, we load capacitors along the *x*-direction of the TL to obtain a frequency-dependent equivalent relative permeability along the *z*-direction. Because the unit size in the TL system is much smaller than the wavelength, the quasi-static, TE-polarized effective permittivities and permeabilities of the 2D TLs can be written as [65, 66]:

$$\varepsilon = 2C_0 g / \varepsilon_0 - g / \omega^2 L d \varepsilon_0, \quad \mu_x = L_0 / g \mu_0, \tag{6}$$
$$\mu_x = L_0 / g \mu_0 - 1 / g \omega^2 C d \mu_0,$$

where $\varepsilon_0$ and $\mu_0$ are the permittivity and permeability of vacuum, respectively.

$C_0$ and $L_0$ denote the capacitance and inductance of the TL per-unit length, respectively. Using Eq. (6), $\mu_x = 1$ (dashdotted) and the dependence of $\mu_z$ (green solid line) and $\varepsilon$ (red dashed line) on the frequency are drawn in Fig. 6(c). At the critical frequency 1.2 GHz, the effective parameters are $\mu_x = 1$, $\mu_z = -0.3$, and $\varepsilon \to 0$, which corresponds to an NLCMM [34, 35]. After rotating the anisotropic circuit-based media, the effective EM parameters are calculated using Eq. (3). In this case, the dispersion relation in the TLs can be expressed as

$$Ak_x^2 + Bk_z^2 + Ck_x k_z = \varepsilon \mu_x \mu_z k_0^2, \tag{7}$$

where $A = \mu_x \cos^2(\alpha) + \mu_z \sin^2(\alpha)$, $B = \mu_x \sin^2(\alpha) + \mu_z \cos^2(\alpha)$, and $C = \sin(2\alpha)(\mu_x - \mu_z)$. Three IFCs of TLCMMs with different frequencies are used as examples in Fig. 6(d). Specifically, the critical case of the TLCMM with $\theta_c = \alpha$ appears when the frequency is 1.2 GHz. The DPS medium realized by the circuit-based structure is an isotropic medium ($\varepsilon \approx 3.63$, $\mu = 1$), with an IFC corresponding to a closed circle.

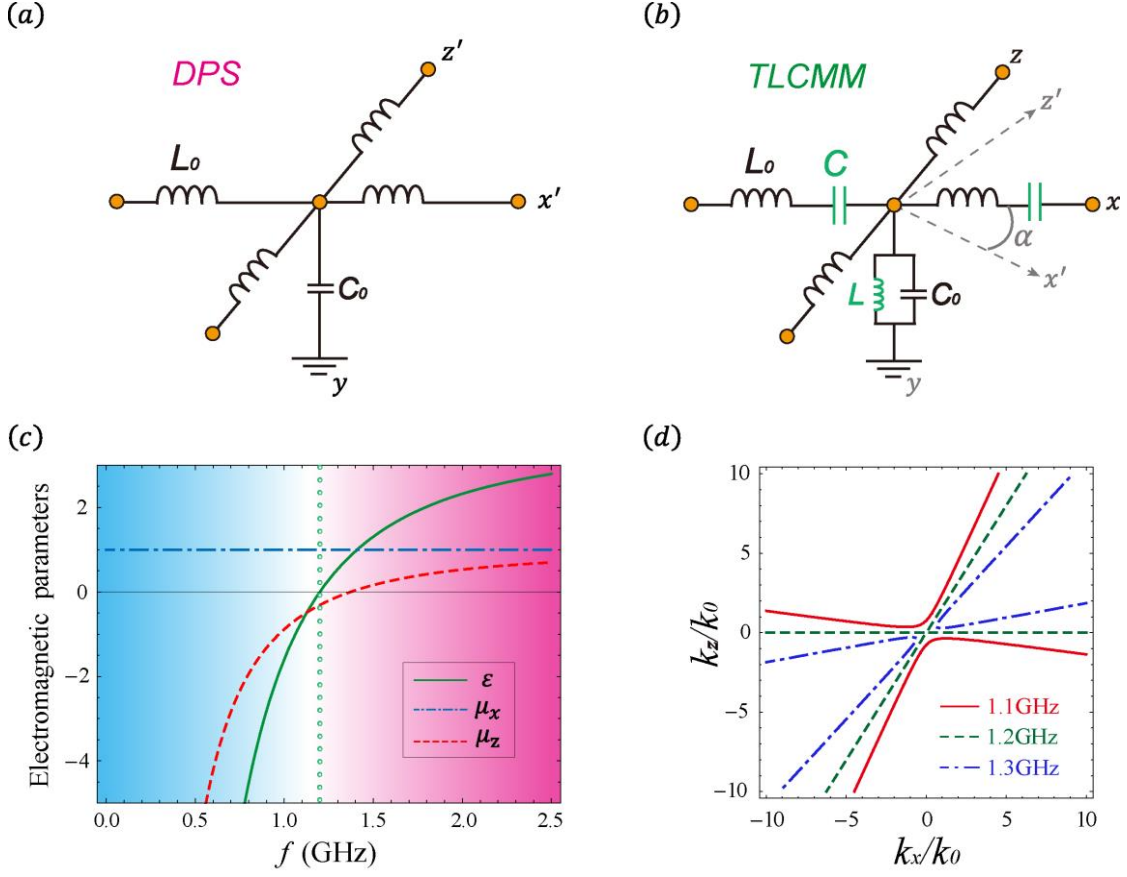

**FIG. 6. Experimental schematic of the circuit-based TLCMM.** (a) A 2D circuit model of the DPS medium. (b) A 2D circuit model of the TLCMM. (c) The effective EM parameters based on the TLs with lumped capacitors and inductors. (d) Topological transition of the tilted anisotropic system. The IFC of the TLCMM is shown by the green, dashed line at 1.2 GHz. For comparison, the IFCs of two THMMs at 1.1 GHz and 1.3 GHz are given by the solid red and dashdotted blue lines, respectively.

Next, we study abnormal negative refraction without reflection and the spatial filtering in the TLCMMs. According to Fig. 3(f), the first sample is composed of the TLCMM (the left part) and a (DPS medium (the right part). In the experimental process, signals were generated by a vector network analyzer (Agilent PNA Network Analyzer N5222A). A subminiature version A (SMA) connector that functioned as the source for the system was placed near the edge of the sample as a vertical monopole to excite the circuit-based prototype. A small, 2 mm-long homemade rod antenna was employed to measure the out-of-plane electric field $E_y$ at a fixed height of 1 mm from the planar microstrip. The circuit-based TLCMMs are constructed by loading lumped elements into the 2D TLs, as shown in Figs. 7(a) and 7(b). When the high-$k$ mode is excited in the TLCMM, it naturally refracts to the DPS medium with a negative refractive angle. The measured normalized $E_y$ distribution for the abnormal refraction without reflection is shown in Fig. 7(c). The measured and simulated results are similar. However, when the source is placed in the DPS medium instead of the TLCMM, refraction of the EM wave changes significantly when it passes through the DPS-TLCMM interface. The EM-wave spatial filtering can be achieved using TLCMMs with $\alpha = -61^o$. Specially, the EM waves are inhibited in the lower half of the structure, as shown in Fig. 7(d).

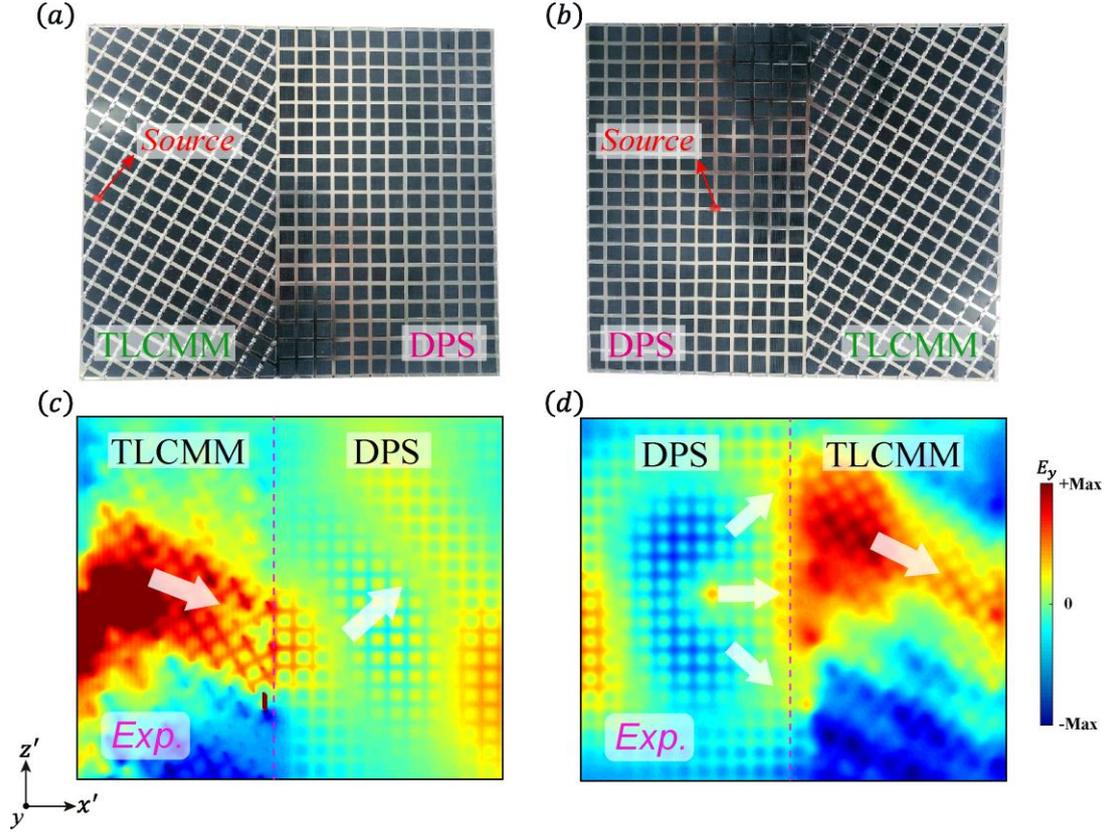

**FIG. 7. Demonstration of abnormal negative refraction without reflection and filtering by circuit-based TLCMMs.** (a) (b) Photos of the TL-based samples. The source are marked by the red arrows. (c) (d) The measured normalized $E_y$ patterns correspond to (a) (b) at 1.2 GHz are given. The directions of the incident and refracted EM waves are marked by white arrows. The interface between the TLCMM and the DPS medium is marked by the pink dashed line.

## 4. Conclusion

In summary, we treated rotation of the optical axis in anisotropic materials as a degree of freedom in order to propose the TLCMM. We went beyond beam splitting and focusing of an NLCMM to demonstrate abnormal negative refraction without reflection of high-*k* modes theoretically and experimentally. Moreover, we discovered that the TLCMM may be used for shielding and filtering. We also studied directional transmission and spatial filtering in heterojunction structures composed of two types of TLCMMs. Our results not only provide a versatile platform to study the abundant and novel IFC, but also may be useful in various planar integrated microwave photonics applications, such as unidirectional power transfer, switching, shielding, and filters.

## ACKNOWLEDGMENTS

This work was supported by the National Key R&D Program of China (Grant No. 2016YFA0301101), the National Natural Science Foundation of China (NSFC) (Grant Nos. 11774261, 11474220, and 61621001), the Natural Science Foundation of Shanghai (Grant Nos. 17ZR1443800 and 18JC1410900), the Shanghai Super Postdoctoral Incentive Program, and the China Postdoctoral Science Foundation (Grant Nos. 2019TQ0232 and 2019M661605).